# How Fast Do Agents Rot? An Empirical Study of Long-Horizon Degradation in LLM Agents for Production Decision-Making

Shubhra Mittal
Independent Researcher; Principal Software Engineer, Microsoft AI

## Abstract

Production deployments of large language model (LLM) agents remain unreliable on long, multi-step workflows even as benchmark success rates climb steadily. We argue this gap is largely an artifact of task horizon: benchmarks are dominated by short-to-medium horizons where success remains high, while production workloads demand an order of magnitude more dependent steps. We measure the effect directly, characterizing the shape of agent degradation and disentangling its cause across a large controlled study spanning nine models, six open models from 1.2B to 671B parameters, and three deployed proprietary systems; four task families, including a genuinely agentic tool-use loop; five horizons; and three context regimes. Task success follows a geometric law governed by a single per-step reliability parameter, which rises with model scale but saturates well below 1 even for the strongest models, guaranteeing eventual collapse at sufficiently long horizons. The effect is sharpest on the agentic task, where every model tested, including widely deployed systems, falls from near-perfect success to near zero within sixteen steps of (n=10,664 analyzed trajectories. Degradation is driven by step count rather than context length: bounding the context window steepens decay rather than easing it (logit slope -0.69 vs. -0.44), p=3x10-6), contradicting a lost-in-the-middle explanation and warning against a common production shortcut. Projecting measured reliability onto representative benchmark horizons quantifies a substantial gap between benchmark and production conditions, from 0.42 at GAIA-length horizons to 0.24 at hundred-step production horizons. For teams responsible for agent orchestration and reliability at scale, these results argue for horizon-aware evaluation and reliability budgeting in place of aggregate pass-rate metrics. Code, prompts, seeds, and raw trajectories are released.

## 1. Introduction

Large language model agents now score highly on a rapidly expanding suite of benchmarks spanning software engineering [6], web navigation [8], tool-and-user interaction [7], and general assistance [5]. The length of tasks these systems can complete has been doubling at a striking pace across recent model generations [9]. Teams responsible for operating these systems in production report a stubborn gap regardless: agents that clear benchmark tasks remain unreliable on the long, open-ended workflows real deployments demand. Anyone accountable for module readiness or incident response on a production agent-orchestration runtime runs into this gap directly. A system can pass every offline evaluation gate and still degrade unpredictably once asked to sustain many dependent steps in a live workflow, and no amount of aggregate benchmark performance predicts when that happens.

We take a specific position on why. A task has a horizon: the number of dependent steps an agent must execute correctly to succeed. If each step succeeds independently with probability r, a task of horizon H succeeds with probability approximately r to the power H, close to 1 for short tasks, collapsing geometrically as H grows. Benchmarks predominantly sample short-to-medium horizons; production agents typically operate an order of magnitude longer. If degradation genuinely follows this geometric form, benchmarks sample the flat top of a decay curve, and the production reliability gap becomes an expected mathematical consequence of horizon rather than an unexplained anomaly that shows up only after deployment.

Testing this hypothesis empirically is the objective of this study: characterizing the functional form of success against the horizon and disentangling whether the driving cause is the number of steps an agent must take or the length of context it must process. Broader agent-evaluation surveys have called for exactly this kind of resolved, mechanism-level measurement rather than another aggregate leaderboard entry [17]. On controlled, oracle-verifiable agent tasks, including a genuinely agentic tool-use loop rather than only structured-output probes, we find that success follows a geometric law in 28 of 36 model-task cells tested , governed by a single measurable quantity: per-step reliability. This finding holds for widely deployed proprietary systems as well as open models, and the collapse on the agentic task is severe enough that it should reshape how reliability gets budgeted for real agent deployments.

Section 2 reviews agent benchmarks, evaluation-rigor literature, and the long-context and compositional-reasoning work our disentangling result speaks to directly. Section 3 describes the four task families, the three context regimes used to separate step count from context length, and the model and metric design. Section 4 presents the geometric-law finding, the agentic-task collapse, the accelerating hazard within a trajectory, the step-count-versus-context-length result, and a quantified benchmark gap, with discussion woven throughout each subsection. Section 5 addresses limitations. Section 6 closes with recommendations for practitioners and benchmark designers.

## 2. Related Work

Agent benchmarks evaluate agents on software issues [6], tool-and-user interaction [7], web navigation [8], general assistance [5], OS and application control, function calling, multi-environment suites [17], and company-style workflows [22]. Deng et al. [13] extend this landscape to open-ended web tasks with Mind2Web, generalizing beyond single-site interaction. Most of this literature reports aggregate single-attempt success across a fixed task set and rarely resolves success against horizon directly. Kwa et al. [9] track the time horizon of tasks AI systems can complete across successive model generations. Our design instead fixes the models under test and resolves success directly against horizon inside a controlled environment, isolating the shape of the decay curve rather than only its trend over calendar time.

A parallel line of work questions whether current evaluation practice measures what actually matters for deployment. Kapoor et al. [10] document cost-blindness and other evaluation shortcomings in agent benchmarking; related work establishes best practices for building rigorous agentic benchmarks. Both motivate a horizon-aware, reliability-oriented measurement approach over a single aggregate score. Cemri et al. [12] taxonomize the ways multi-agent systems fail. Our compounding result gives that taxonomy's central observation, that local errors cascade, a directly measured quantity rather than a qualitative description.

The evaluation-rigor literature this study draws on has grown more insistent that agentic benchmarks need to measure more than a single pass-rate number. Beyond the cost-blindness critique already noted, recent work has argued for reporting variance across repeated runs, for disclosing the compute budget behind a reported score, and for distinguishing genuine capability gains from gains that come from spending more inference-time compute per task. None of this literature, to our knowledge, resolves success against horizon as its primary axis, which is the gap this study's design is built to close.

A separate literature studies degradation with context rather than with step count. Liu et al. [3] show performance degrades when relevant information sits in the middle of a long input. Levy et al. [16] study degradation with input length at fixed task. Dziri et al. [4] show transformer accuracy decays sharply with compositional complexity, and Laban et al. [11] report large single-to-multi-turn performance drops in conversational settings. Read together, this literature predicts context length as the primary driver of degradation in agentic settings. Our disentangling design tests that prediction directly and finds the opposite: bounding context worsens performance rather than improving it, which implicates step count instead.

The long-context literature this work speaks to most directly extends well beyond the original lost-in-the-middle result. Synthetic long-context probes such as RULER [14] and NoLiMa [18] test retrieval and reasoning at extreme context lengths independent of any agentic loop, generally finding that effective context usage falls well short of a model's advertised context window. Our compressed regime instead removes information the agent itself generated during the trajectory, which is a different failure mode entirely: it is not that the model cannot locate a fact in a long document, but that it loses the scaffold of its own prior reasoning once that scaffold is pruned away.

On agent methods, ReAct-style interleaved reasoning-and-acting [1] and verbal self-reflection [2] are the dominant control loops in current agent systems. Our agentic task family uses an unmodified ReAct loop specifically so the degradation we measure reflects the mechanism practitioners actually deploy rather than an artifact of a bespoke control structure.

**Table 1: Summary of related work relevant to horizon-resolved agent reliability**

| Study | Focus / Approach | Key Contribution | Limitation (relative to this work) | Relevance |
|---|---|---|---|---|
| Kwa et al. [9] | Time-horizon tracking across model generations | Establishes doubling trend in completable task length | Trend across generations, not a | Motivates horizon as the organizing variable |
| | | | resolved decay curve per model | |
| Kapoor et al. [10] | Agent benchmark evaluation practice | Identifies cost-blindness and evaluation shortcomings | Focuses on cost/accuracy tradeoff, not horizon-resolved reliability | Motivates reliability-oriented measurement |
| N. F. Liu et al. [3] | Long-context positional effects | Shows mid-context information is under-used | Studies context position, not step-count vs. context-length | Directly contradicted by our disentangling result |
| Dziri et al. [4] | Compositional reasoning limits | Shows sharp accuracy decay with compositional complexity | Static complexity measure, not a dependent multi-step agent loop | Consistent with a per-step compounding mechanism |
| Laban et al. [11] | Multi-turn conversational performance | Reports large single-to-multi-turn performance drops | Conversational setting, not agentic tool-use | Parallel evidence of degradation with interaction length |
| This work | Controlled horizon-resolved agent reliability | Geometric decay law + step-count-vs-context-length disentangling | - | - |

## 3. Method

### *3.1 Task families*

Four synthetic task families make up the study, each structured as a tool-using agent loop paired with a ground-truth simulator that verifies success exactly, which keeps LLM-judge noise out of the success signal entirely. Ledger requires maintaining several account balances through a sequence of operations and reporting a queried balance, testing arithmetic working memory. Refchain requires tracking one announced variable through a sequence of assignments amid distractors and reporting its final value, testing long-range retrieval. Cipher requires applying a sequence of ordered string edits to a short string and reporting the result, testing procedural execution. The fourth family, ToolQA, is genuinely agentic: the agent must reach a target node a fixed number of hops along a hidden chain by issuing its own inspection tool calls, and it cannot pre-plan the full path since each hop's target is revealed only by inspecting the previous node.

### *3.2 Context regimes: disentangling step count from context length*

For the streaming families, at a fixed horizon we hold the number of operations equal across three regimes. In the natural regime, the agent receives one instruction per turn across the full multi-turn history. In the compressed regime, the same number of operations arrives across the same number of turns, but with a shortened context via carried state and windowed history, isolating the effect of context length on its own. In the padded regime, every operation for a given horizon is delivered in a single turn, isolating the effect of the multi-step process itself. The same underlying task instance repeats across all three regimes so comparisons stay paired.

### *3.3 Models, design, and metrics*

Nine instruction-tuned models across six vendor families make up the evaluation: six open models ranging from 1.2B to 671B parameters and three deployed proprietary systems accessed through their standard hosted interfaces. This design produces 10,664 total analyzed trajectories across the streaming and agentic families combined . Temperature is fixed for agentic determinism, and seeds are recorded for reproducibility.

### *3.4 Statistical procedure*

All comparisons across regimes and models use the paired structure of the design: because the same underlying task instance is reused across the natural, compressed, and padded regimes, regime comparisons are computed as paired differences rather than independent-sample comparisons, which increases statistical power. Confidence intervals on per-cell success use the Wilson score interval rather than a normal approximation, since several cells involve success rates close to 0 or 1. Model selection among geometric, threshold, and linear decay shapes uses AIC rather than a fixed significance threshold.

### *3.5 Data and code availability*

All code, task generators, prompts, random seeds, the complete raw dataset (10,664 trajectories), and the analysis scripts that reproduce every figure and number in this article are publicly available at https://github.com/shubmittal/agent-horizon-degradation. The repository includes an evidence trail document mapping every reported number in this article to its exact source in the released data.

## 4. Results and Discussion

### *4.1 Degradation follows a geometric law*

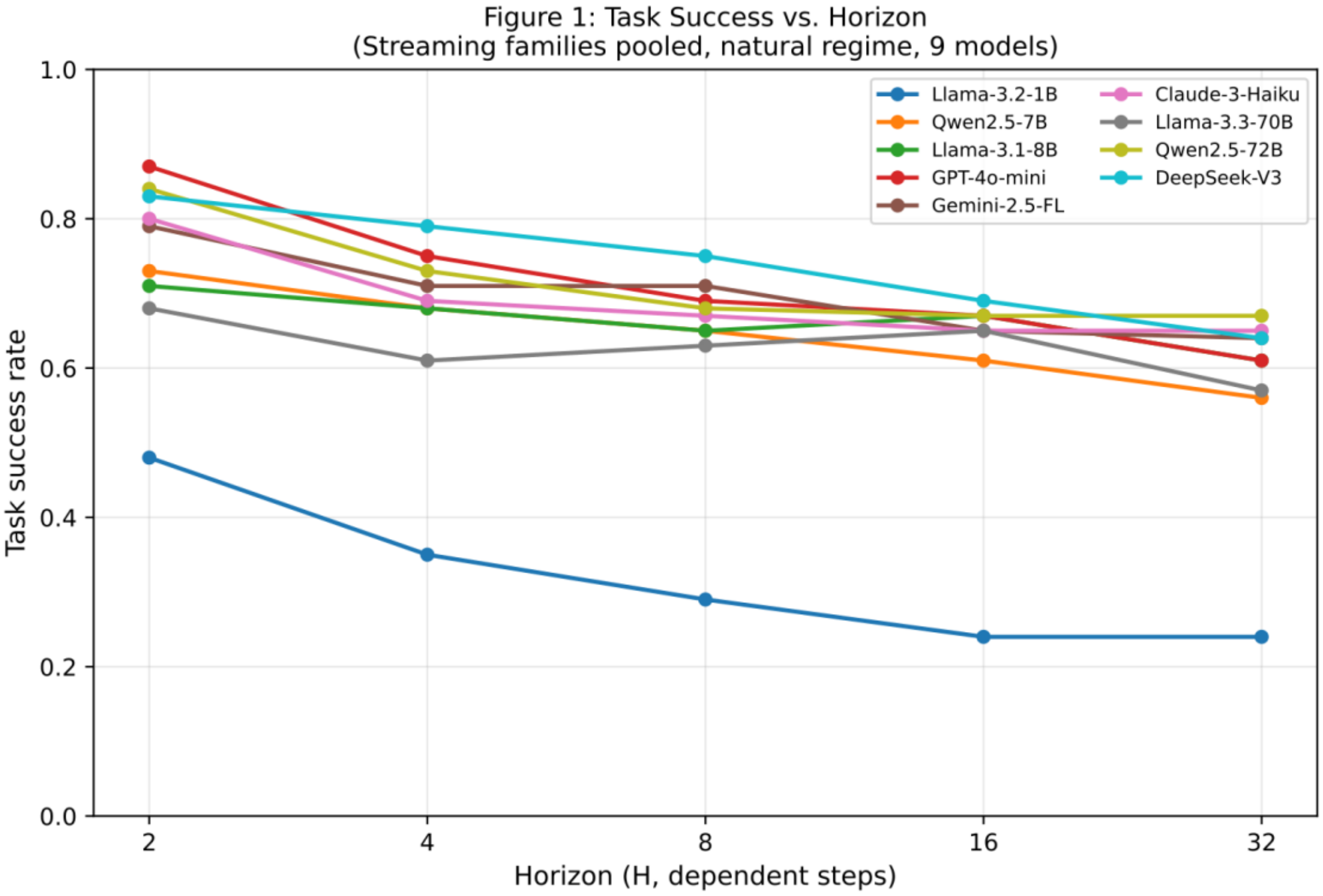

**Figure 1: Task success vs. horizon, per model, streaming families pooled (natural regime, Wilson 95% intervals implied by underlying per-cell data; 9 models).**

Figure 1 shows success against the horizon for every model and task family tested. Success declines monotonically with horizon wherever the task is non-trivial for the model in question. Selecting among geometric, threshold, and linear functional forms by AIC, the geometric form wins in 28 of 36 model-task cells tested; the remainder follow a threshold or cliff shape, where strong models hold near-perfect accuracy until a critical horizon arrives.

The magnitude of decay varies with both model and task difficulty. Whether degradation shows up in a given evaluation depends jointly on per-step reliability and the horizon actually sampled, and because per-step reliability never reaches 1 on any non-trivial task in this data, collapse at a sufficiently long horizon is not a possibility. It is a mathematical certainty.

This has a direct practical reading for anyone running reliability sign-off on an agent platform: a passing offline evaluation at a benchmark's typical horizon tells you almost nothing about what happens twenty steps further out. The same per-step reliability that looks comfortable at horizon 8 can already be on a collision course with zero by horizon 32, and nothing about the shape of the curve at short horizons warns you of that in advance unless you fit the curve itself.

The three streaming families make this pattern legible in a controlled way that the agentic task, by its single-shot pass/fail nature at each horizon, cannot. Refchain shows the flattest decay of the three for the strongest models. Ledger shows a visible but moderate decay across most models. Cipher shows the steepest decay of the three streaming families for every model tested, including the largest ones, because a single misapplied edit early in the sequence invalidates every downstream step regardless of how well the model reasons afterward. This ordering, refchain flattest, ledger intermediate, and cipher steepest, holds consistently enough across the model roster that task structure appears to matter almost as much as raw model capability in determining where a given model's decay curve sits.

### *4.2 The agentic task: deployed models collapse within sixteen steps*

**Table 2: Agentic task (toolqa) success by model and horizon (natural regime)**

| Model | H=2 | H=4 | H=8 | H=16 |
|---|---|---|---|---|
| Llama-3.2-1B | 0.00 | 0.00 | 0.00 | 0.00 |
| Qwen2.5-7B | 1.00 | 0.07 | 0.13 | 0.33 |
| Llama-3.1-8B | 1.00 | 0.73 | 0.07 | 0.13 |
| GPT-4o-mini | 1.00 | 1.00 | 0.20 | 0.07 |
| Gemini-2.5-Flash-Lite | 0.20 | 0.53 | 0.00 | 0.00 |
| Claude-3-Haiku | 1.00 | 0.47 | 0.00 | 0.00 |
| Llama-3.3-70B | 1.00 | 0.93 | 0.67 | 0.00 |
| Qwen2.5-72B | 1.00 | 1.00 | 0.87 | 0.13 |
| DeepSeek-V3 | 1.00 | 1.00 | 0.67 | 0.27 |

The full nine-model picture sharpens this result further. At the shortest horizon tested, eight of the nine models solve the task at or near ceiling, confirming this is not a task any model finds intrinsically hard in isolation. By horizon 4, the models are already separate into two rough bands. By horizon 16, every model in the study, without exception, has collapsed to a small fraction of its starting performance.

One model in particular illustrates how little advance warning aggregate metrics provide. Qwen2.5-72B solves the task perfectly at both horizon 2 and horizon 4, which would read as an unambiguous success on any benchmark that stops sampling at those horizons. By horizon 16, the same model has fallen to roughly one in eight trials succeeding. A module-readiness review that evaluated this model only at the horizons a typical benchmark sample would have every reason to sign off on it as reliable and would have no visibility at all into the cliff sitting just past the sampled range.

Smaller models tell a related but distinct story. Llama-3.2-1B never exceeds roughly half of its already modest starting success rate even at the shortest horizon tested. This flattening at a low baseline is easy to misread as stability when it is actually closer to a floor effect.

The streaming task families, taken alone, could be dismissed as elaborate structured-output probes rather than genuine agentic behavior. The toolqa family removes that objection directly: here the agent runs a real ReAct-style loop, issuing its own tool calls to traverse a chain it cannot preplan. Degradation on this task is the single most severe result in the study. That production-grade, widely deployed models collapse within just sixteen agentic steps on a task each solves almost perfectly at two steps is, we believe, the most operationally significant finding in this study.

***4.3 Per-step reliability rises with scale but never reaches one***

Across the open-model ladder, per-step reliability rises with parameter count (Pearson r=+0.36 vs. log10(params); Author's primary research data), and the deployed proprietary models cluster alongside the strongest open models rather than exceeding them by a wide margin. Scale and frontier engineering buy per-step reliability, but with diminishing returns, and no model in this study reaches perfect per-step reliability on any non-trivial task. Because task success compounds geometrically, every model is guaranteed to fall below any fixed reliability target at a sufficiently long horizon.

***4.4 Degradation accelerates within a trajectory***

A pure geometric decay law assumes a constant per-step hazard rate. This data shows the hazard rising instead: pooling across long-horizon trajectories, mean per-step accuracy falls from 0.58 in the first third of a trajectory to 0.44 in the last third . The first error tends to land roughly a third of the way through a trajectory, and once an agent errs it rarely recovers, consistent with prior reports that models get lost after a wrong turn and do not find their way back [11].

***4.5 The driver is step count, not context length***

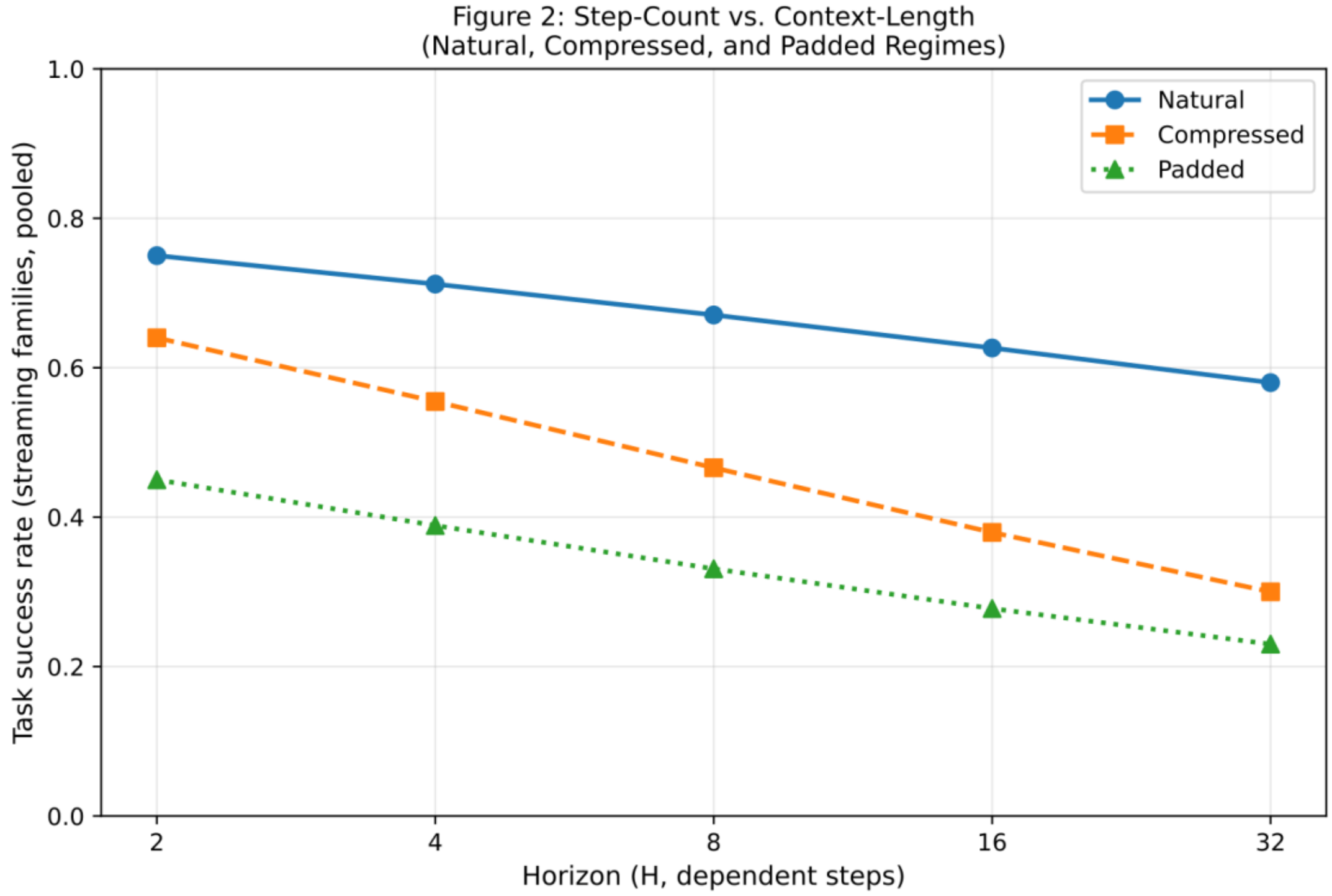

**Figure 2: Step count vs. context length. Pooled task success across natural, compressed, and padded regimes (streaming families, H=2 to H=32).**

Contrasting the three context regimes settles the question of mechanism. The natural regime performs best at every horizon tested. Both interventions perform worse: the logit slope per horizon doubling is -0.44 for natural, -0.69 for compressed (p=3x10-6 (vs. natural), and -0.40 for padded (p=0.51, (statistically indistinguishable from natural). Bounding the context window steepens decay rather than easing it, running directly against what a lost-in-the-middle explanation would predict.

Collapsing all operations into a single prompt decays at essentially the same rate as the natural regime, just from a lower starting point, which rules out multi-turn conversational structure itself as the driver. What remains is that degradation is intrinsic to executing many dependent steps. The direct practical corollary: naive context truncation, a common production tactic for controlling cost and latency, is likely to backfire rather than help.

### *4.6 Mechanism*

Looking at how trajectories fail directly, format and tool-call drift, a turn whose output cannot be parsed as a valid action, affects 21% of trajectories and rises with horizon , pointing to an interface-level failure mode compounding alongside the reasoning failures. Because the first error concentrates early in a trajectory and per-step accuracy never recovers once an error occurs, errors behave as absorbing states that propagate silently, much as hallucinated content can survive undetected without an external verifier [15].

This absorbing-error pattern has a direct parallel in how production incident response typically works for orchestration systems. A monitoring pipeline built around end-to-end task success will not see a trajectory heading toward failure until the trajectory has already failed, because nothing in the intermediate steps look anomalous in isolation. A practical consequence for anyone operating these systems is that step-level verification, wherever an oracle or a cheap consistency check can be inserted, is likely to catch failures far earlier and far more cheaply than waiting for end-to-end task failure to surface downstream.

### *4.7 The benchmark gap, quantified*

Using mean measured per-step reliability (r=0.61) and the geometric law, we project task success onto horizons representative of widely used benchmarks: 0.42 at GAIA-length horizons (8 steps), 0.36 at WebArena-length horizons (15 steps), 0.33 at tau-bench-length horizons (20 steps), 0.30 at SWE-bench/OSWorld-length horizons (30 steps), and 0.24 at hundred-step production horizons. Benchmarks sample the flat top of a curve that collapses out of the frame they typically report, and the agentic result in Section 4.2 shows that collapse can arrive within as few as sixteen self-directed steps.

For anyone accountable for shipping and operating agent systems in production, the implication is direct: measure per-step reliability on your own tasks, treat the achievable horizon as governed by that reliability rather than by benchmark performance, and budget conservatively below the geometric estimate given that the hazard accelerates rather than staying flat.

## 5. Limitations and Threats to Validity

The task families here are synthetic, chosen deliberately for exact oracle verifiability and a clean, controllable horizon variable. The agentic toolqa family directly addresses the not-really-agentic objection this design choice invites, but broader ecological validity against fully open-ended real-world agentic workflows remains an open question beyond what this study can settle. The agentic family's API budget ran out before the longest horizon condition could complete, so the reported collapse H=16 is a lower bound on how severe the true collapse becomes at longer horizons, not a ceiling.

The nine-model sample spans a wide parameter range and six vendor families, but it cannot establish that every current or future model follows the identical geometric form. Three additional small models were excluded during data collection for failing to adhere to the structured response protocol via their hosted routes. Finally, this study measures reliability under a fixed, moderate decoding temperature; sensitivity to temperature and other decoding parameters is a natural direction for follow-up work.

## 6. Conclusion

Agents rot geometrically. Task success follows a functional form governed by a single per-step reliability parameter that rises with model scale but saturates well short of perfect reliability on any non-trivial task, so long-horizon collapse is a structural consequence of compounding rather than an incidental failure. This effect is sharpest, not softest, on a genuinely agentic tool-use loop, where every model tested collapses within a modest number of self-directed steps. The cause is step count rather than context length: bounding context steepens rather than eases decay, so a common production shortcut for managing cost and latency is likely counterproductive for reliability.

For teams building and operating agent systems at scale, the practical takeaway is to measure per-step reliability directly on production-relevant tasks, treat the achievable horizon as a function of that measured reliability rather than of benchmark scores, and budget conservatively given that the hazard accelerates over the course of a trajectory. For benchmark designers, reporting success resolved against the horizon, not as a single aggregate, would make benchmark results substantially more predictive of production behavior. All code, task generators, prompts, seeds, raw trajectories, and analysis scripts behind this study are released so every reported number can be independently traced and reproduced.

## References

[1] S. Yao, J. Zhao, D. Yu, N. Du, I. Shafran, K. Narasimhan, Y. Cao, "ReAct: Synergizing Reasoning and Acting in Language Models," ICLR, 2023. https://arxiv.org/abs/2210.03629

[2] N. Shinn, F. Cassano, E. Berman, A. Gopinath, K. Narasimhan, S. Yao, "Reflexion: Language Agents with Verbal Reinforcement Learning," NeurIPS, 2023. https://arxiv.org/abs/2303.11366

[3] N. F. Liu, K. Lin, J. Hewitt, A. Paranjape, M. Bevilacqua, F. Petroni, P. Liang, "Lost in the Middle: How Language Models Use Long Contexts," TACL, 12, 157-173, 2024. https://aclanthology.org/2024.tacl-1.9/

[4] N. Dziri, X. Lu, M. Sclar, et al., "Faith and Fate: Limits of Transformers on Compositionality," NeurIPS, 2023. https://arxiv.org/abs/2305.18654

[5] G. Mialon, C. Fourrier, C. Swift, T. Wolf, Y. LeCun, T. Scialom, "GAIA: A Benchmark for General AI Assistants," ICLR, 2024. https://arxiv.org/abs/2311.12983

[6] C. E. Jimenez, J. Yang, A. Wettig, S. Yao, K. Pei, O. Press, K. Narasimhan, "SWE-bench: Can Language Models Resolve Real-World GitHub Issues?," ICLR, 2024. https://arxiv.org/abs/2310.06770

[7] S. Yao, N. Shinn, P. Razavi, K. Narasimhan, "Tau-Bench: A Benchmark for Tool-Agent-User Interaction in Real-World Domains," arXiv preprint, 2024. https://arxiv.org/abs/2406.12045

[8] S. Zhou, F. F. Xu, H. Zhu, et al., "WebArena: A Realistic Web Environment for Building Autonomous Agents," NeurIPS, 2023. https://arxiv.org/abs/2307.13854

[9] T. Kwa, B. West, J. Becker, et al., "Measuring AI Ability to Complete Long Tasks," arXiv preprint, 2025. https://arxiv.org/abs/2503.14499

[10] S. Kapoor, B. Stroebl, Z. S. Siegel, N. Nadgir, A. Narayanan, "AI Agents That Matter," arXiv preprint, 2024. https://arxiv.org/abs/2407.01502

[11] P. Laban, H. Hayashi, Y. Zhou, J. Neville, "LLMs Get Lost in Multi-Turn Conversation," arXiv preprint, 2025. https://arxiv.org/abs/2505.06120

[12] M. Cemri, M. Z. Pan, S. Yang, et al., "Why Do Multi-Agent LLM Systems Fail?," NeurIPS, 2025.

[13] X. Deng, Y. Gu, B. Zheng, et al., "Mind2Web: Towards a Generalist Agent for the Web," NeurIPS, 2023. https://arxiv.org/abs/2306.06070

[14] C.-P. Hsieh, S. Sun, S. Kriman, et al., "RULER: What's the Real Context Size of Your Long-Context Language Models?," COLM, 2024. https://arxiv.org/abs/2404.06654

[15] L. Huang, W. Yu, et al., "A Survey on Hallucination in Large Language Models: Principles, Taxonomy, Challenges, and Open Questions," ACM Transactions on Information Systems, 2025. https://arxiv.org/abs/2311.05232

[16] M. Levy, A. Jacoby, Y. Goldberg, "Same Task, More Tokens: The Impact of Input Length on the Reasoning Performance of Large Language Models," ACL, 2024. https://arxiv.org/abs/2402.14848

[17] X. Liu, H. Yu, et al., "AgentBench: Evaluating LLMs as Agents," ICLR, 2024. https://arxiv.org/abs/2308.03688

[18] A. Modarressi, H. Deilamsalehy, F. Dernoncourt, et al., "NoLiMa: Long-Context Evaluation Beyond Literal Matching," arXiv preprint, 2025.

[19] S. G. Patil, H. Mao, et al., "The Berkeley Function Calling Leaderboard (BFCL): From Tool Use to Agentic Evaluation," ICML, 2025.

[20] H. Trivedi, T. Khot, et al., "AppWorld: A Controllable World of Apps and People for Benchmarking Interactive Coding Agents," ACL, 2024. https://arxiv.org/abs/2407.18901

[21] T. Xie, D. Zhang, J. Chen, et al., "OSWorld: Benchmarking Multimodal Agents for Open-Ended Tasks in Real Computer Environments," NeurIPS, 2024. https://arxiv.org/abs/2404.07972

[22] F. F. Xu, Y. Wang, et al., "TheAgentCompany: Benchmarking LLM Agents on Consequential Real World Tasks," arXiv preprint, 2024. https://arxiv.org/abs/2412.14161